\documentclass{sig-alternate}
\usepackage{amsmath}
\usepackage{txfonts}
\usepackage{amssymb}
\usepackage{graphicx}
\usepackage{booktabs}
\usepackage{comment}
\usepackage{subfigure}

\newtheorem{theorem}{Theorem}

\newtheorem{problem}{Problem}

\newtheorem{property}{Property}
\newtheorem{assumption}{Assumption}

\def\newblock{\hskip .11em plus .33em minus .07em}
\DeclareGraphicsExtensions{.pdf,.png,.jpg,.eps}
\usepackage[lined,ruled,vlined]{algorithm2e}

\begin{document}

  \title{A Social Influence Model Based On Circuit Theory}
  \numberofauthors{1}
\author{
\alignauthor{Biao Xiang$^1$, Enhong Chen$^1$, Qi Liu$^1$, Hui Xiong$^2$, Yu Yang$^1$, Junyuan Xie$^1$}\\
\vspace{0.1cm}
\affaddr{$^1$Computer Science and Technology Department, University of Science and Technology of China }\\
\affaddr{E-mail: cheneh@ustc.edu.cn, [bxiang, feiniaol, ryanyang, dinic]@mail.ustc.edu.cn}\\
\affaddr{\vspace{0.1cm}$^2$ MSIS Department, Rutgers University, E-mail: hxiong@rutgers.edu}\\
}
\maketitle
\begin{abstract}
Understanding the behaviors of information propagation is essential for the effective exploitation of social influence in social networks. However,  few existing influence models are tractable and efficient for describing the information propagation process, especially when dealing with the difficulty of incorporating the effects of combined influences from multiple nodes. To this end, in this paper, we provide a social influence model that alleviates this obstacle based on electrical circuit theory. This model vastly improves the efficiency of measuring the influence strength between any pair of nodes, and can be used to interpret the real-world influence propagation process in a coherent way. In addition, this circuit theory model provides a natural solution to the social influence maximization problem. When applied to real-world data, the circuit theory model consistently outperforms the state-of-the-art methods and can greatly alleviate the computation burden of the influence maximization problem.
\end{abstract}

\section{Introduction}

A social network is a graph of relationships between individuals,
groups, or organizations. As an effective tool in connecting people
and spreading information, social networks have caught the attention
of millions of small businesses as well as major companies. To
exploit social networks to gain influence, it is essential to
understand the behaviors of information propagation in social
networks. However, given the large-scale of existing social
networks, it is challenging to design a tractable and efficient
model to describe the information propagation process in social
networks. For instance, let us consider a scenario that an event
(such as a rumor) initially happened on a node (denoted
as the seed node), this event may happen again on some neighboring
nodes and may further happen on some far-away nodes due to the
information propagation through the network. Then, the question is
how does the information propagate? Also, after sufficient
propagation, what probability will this event happen on a random
node? (this probability could be viewed as the \textbf{influence}
strength from the seed node to this random node.)

Indeed, there are several existing models to describe the above
information propagation process, such as the Linear Threshold(LT) model \cite{granovetter1978threshold}
the Independent Cascade(IC) model~\cite{goldenberg2001talk}. However, these models are operational models, they are untractable and inefficient. Under these models, if we want to get the probability of one event happened on a node, we need to run Monte-Carlo simulations of these influence models for sufficiently
many times (e.g. 20000 times) to obtain an accurate estimate of the
probability for the event happen~\cite{chen2009efficient}, and this
approach is very time-consuming. Moreover, if the event initially
happened  on more than one node (seed), the probability of this
event happened on a random node will be a combination of influences
from all of these  seeds. In this case, how to get the probability
for the event happen and how to identify the independent probability
(or independent influence) of each seed are challenging research
issues.

To this end, in this paper, we propose a social influence model
based on electrical circuit theory to simulate the information
propagation process in social networks. With this model, we are able
to obtain a probabilistic influence matrix to describe the influence
strength between any pair of nodes. Also, based on this model, we
provide a novel method to identify the independent influence of each
seed node when there are more than one seed. This can lead to a
natural solution to the social influence maximization
problem~\cite{kempe2003maximizing, chen2009efficient,
wang2010community}, which targets on finding a small subset of
influential nodes in a social network  to influence as many nodes as
possible. Specifically, the contributions of this paper can be
summarized as follows.

\begin{itemize}
\item We propose the idea of exploiting the circuit network to simulate the information (or influence) propagation process in  social networks. This simulation is straightforward and is easy to interpret and understand.  To the best of our knowledge, this is the first work that exploits circuit theory for estimating the spread of social influence. What's more, this model is tractable even for more than one seed node. With this model, we could provide a novel way to compute the independent influence of a seed naturally.
\item We identify an upper bound of the node's influence on the network. This upper bound property can help us to select the truly influential nodes and drastically reduce the search space. By exploiting these computational properties, we develop an efficient approximation method to compute the influences and successfully apply this circuit theory based model to solve the social influence maximization problem. Along this line,  we first prove that the influence spread function is also submodular under the circuit model. Then, by the greedy strategy, we design an algorithm to select $K$ influential nodes to maximize their influence. Finally, experimental results show that the circuit theory based methods have the advantages over the state-of-the-art algorithms for the social influence maximization problem in terms of both efficiency as well as the effectiveness.
\end{itemize}

\section{A Social Influence Model based on Circuit Theory}~\label{sec:model}
In this section, we propose a social influence model based on electrical circuit theory.

\subsection{Social Influence Modeling}
Social influence refers to the behavioral change of individuals affected by
others in a network. Social influence is an intuitive and well-accepted phenomenon in social networks~\cite{easley2010networks}. In this section, we will propose a quantitative definition of social influence to measure its amount.

Let $G=(V,E)$ is an information network~\footnote{In an information network, the information will flow along with the direction of an arc. Any network could be an information network. For example, twitter's information network should be its inverse, for the information flows from node $j$ to node $i$ on arc $(i,j)$ in twitter.}, where the node set $V$ includes all of individuals, the arc set $E$ represents all the social connections. Let $F_{i\rightarrow j}$ denote the influence from node $i$ to node $j$, we propose three rules to define the influence between any pair of nodes.
\begin{itemize}
\item[1.] The influence from oneself should be 1, that is $F_{i\rightarrow i}=1$.
\item[2.] The influence could transmit through the network arcs with certain probability.
\item[3.] The influence on one node should be related to the influences on the node's neighbors, suppose node $j$'s neighbor set is  $N_j=\{j_1,j_2,...j_m\}$~(i.e. $\forall k\in N_j,\ (k,j)\in E$),  then
\begin{equation}\label{eq:mix}
  F_{i\rightarrow j} = f_j(t_{j_1j}F_{i\rightarrow j_1},\  t_{j_2j}F_{i\rightarrow j_2},...\ t_{j_mj}F_{i\rightarrow j_m})
\end{equation}
where $t_{kj}$ is the transmission probability on arc $(k,j)$.
\end{itemize}

To the construction modeled by the above three rules, there are two factors could change its shape. The first one is the transmission probabilities on arcs. In this paper, we propose an assumption to confine the probability, that is
\begin{assumption}\label{as:theta}
  The sum of transmission probabilities flowing into one node should be less than or equal to 1. That is,
$$
  \theta_i = \sum_{j=1}^{n}t_{ji} \leq 1\ \ \ \ for\ \ i=1,2...n
$$
or
\begin{equation}
  T'\mathbf{e}=\Theta \leq \mathbf{e}
\end{equation}
where $\Theta=diag(\theta_1,\theta_2,...\theta_n)$ and $T=[t_{ij}]_{n*n}$ is a transmission matrix in which $t_{ij}$ is the transmission probability from node $i$ to node $j$. If $(j,i)\not{\in}A$, then $t_{ji}=0$.
\end{assumption}
Actually, this assumption is used for measuring the amount of information~(e.g., with regard to an event or message) that will be accepted by each node. The corresponding value varies in the range of [0,1], where $0$ stands for the ignorance of the message and $1$ means this node totally believes in it. Notably, in the last section of this paper, we also discuss the way how to break through the confinement of this assumption.

The second one is the way how the influence on one node is related to the influences on its neighbors. In the previous works, Aggarawal et al~\cite{aggarwal2011social} proposed a way to describe the relation among the neighbors' influence, that is
\begin{equation}\label{eq:aggarwal}
  F_{i\rightarrow j} = 1-\Pi_{k\in N_j}(1-t_{kj}F_{i\rightarrow k})
\end{equation}
which claims that the transmitted influence from different neighbors should be independent to each other. This way is a theoretically  reasonable way but it is too complex to get its closed-form solution. So, in this paper, we propose a linear method to define it. That is,
\begin{equation}\label{eq:linear}
  F_{i\rightarrow j} = \frac{1}{1+\lambda_j}\sum_{k\in N_j}{t_{kj}F_{i\rightarrow k}} \ \ \ \ for\ j\neq i
\end{equation}
where $\lambda_j$ locates in the range $(0,+\infty)$ which is a damping coefficient of node $j$ for the influence propagation. The smaller the $\lambda_i$ is (i.e., approaching to 0), the less the information will be blocked by node i. This number varies with respect to the topic that the propagating information belongs to. If node $i$ favors the topic of the propagating information, $\lambda_i$ should approach to 0, otherwise, it should approach to $+\infty$. Thus, this way could be topic sensitive when the damping coefficients on each node are  decided according to their favorite.

\subsection{A Physical Implication For Linear Model}\label{sec:circuit}
\begin{figure}[!h]
\begin{center}
    \makebox{\subfigure[  A sample social network.
    \quad]{\includegraphics[scale=0.35]{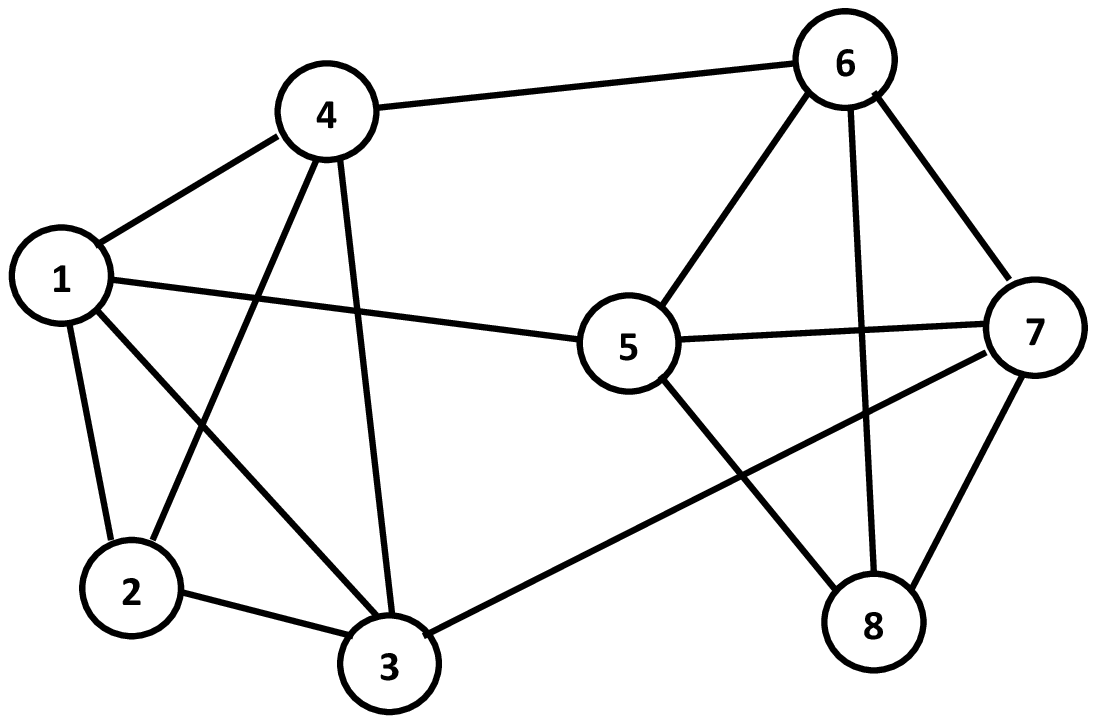}}}   
    \makebox{\subfigure[  The circuit network.
    \quad]{\includegraphics[scale=0.3]{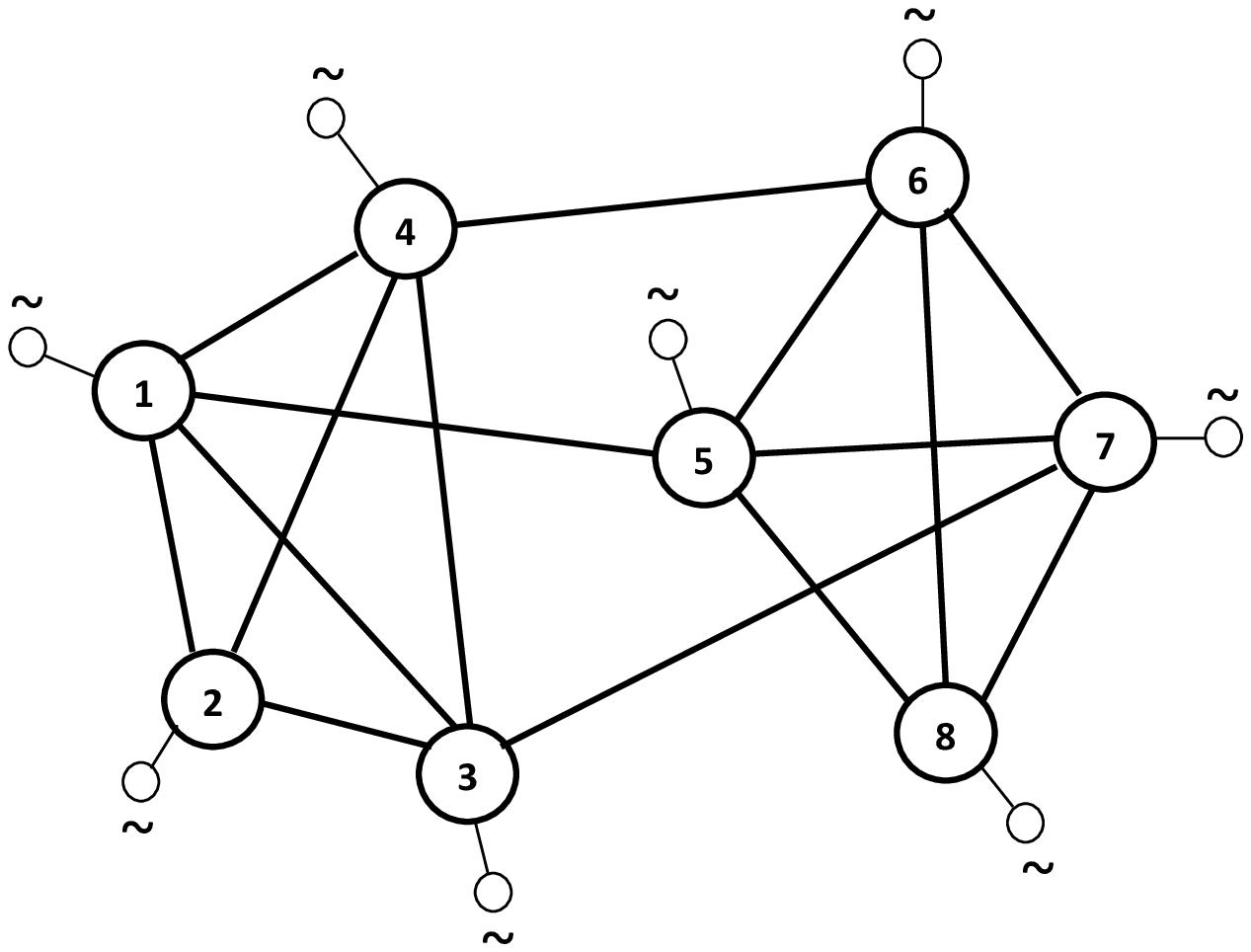}}}   
\end{center} \caption{An Illustration Example.}
\label{fig-network} \vspace{-0.4cm}
\end{figure}

To the linear way proposed in Equation~\ref{eq:linear}, there is a physical implication when the information network $G$ is undirected. We find that when we construct a circuit network $G'$ by the following way, the current flow in the circuit is running in the same way with the information propagation described by Equation~\ref{eq:linear} and the potential value on each node is an equivalent value to the influence on it. For an undirected network $G$ in Figure~\ref{fig-network} (a), we construct its corresponding circuit network $G'$  in Figure~\ref{fig-network} (b) as follows.

\begin{itemize}
  \item First, construct a topologically isomorphic circuit network of $G$, where the conductance between node $i$ and $j$ is equal to $c_{ij}$ and guarantees that $\frac{c_{ij}}{d_j}=\frac{t_{ij}}{\theta_j}$ where $d_j=\sum_{i=1}^{n}{c_{ij}}$;
  \item Second, connect each node $i$ with an external electrode $\tilde{E_i}$ through an additional electric conductor with conductance $\frac{(1+\lambda_i-\theta_i)d_i}{\theta_i}$. The electric potential value on $\tilde{E_i}$ is $\frac{\nu_i}{(1+\lambda_i-\theta_i)}$ where $\nu_i$ is a real number which will be decided later.
\end{itemize}
In $G'$, Kirchhoff equations~\cite{kirchhoff} tell us that the total current flowing
into any node $j$ should sum up to zero, thus in the circuit network
$G^{'}$:
\begin{equation}\label{Kiff1}
\sum{I}_{j}=\sum_{k=1}^{n}{c_{kj}(\tilde{U}_{k}-\tilde{U}_{j})}+\frac{(1+\lambda_j-\theta_j)d_j}{\theta_j}
(\frac{\nu_j}{(1+\lambda_j-\theta_j)}-\tilde{U}_{j})=0
\end{equation}
where $\sum{I}_{j}$ is the total current flowing into node $j$ and $\tilde{U}_{j}$ is the electric potential value on node $j$. From Equation~\ref{Kiff1}, we have
\begin{eqnarray}\label{eq:Kiff2}
\tilde{U}_{j}&=&\frac{1}{d_j+\frac{(1+\lambda_j-\theta_j)d_j}{\theta_j}}(\sum_{k=1}^{n}{c_{kj}\tilde{U}_{k}}+ \frac{\nu_id_j}{\theta_j})\nonumber\\
&=&\frac{1}{1+\lambda_j}(\sum_{k=1}^{n}{t_{kj}\tilde{U}_{k}}+\nu_j)
\end{eqnarray}
In this Equation, if we decide $\nu_j$ by the following way
\begin{equation}
\nu_j=\left\{\begin{aligned}
  &a\ number\ to\ guarantee\ \tilde{U}_{j}=1 & \ \  &j = i& \\
  &0&\ \ \ &j\neq i&
\end{aligned}\right.
\end{equation}
then comparing with Equation~\ref{eq:linear}, it is easy to get
$$
\tilde{U}_{j} = F_{i\rightarrow j}
$$
which means that the potential value on node $j$ is an equivalent value to the influence $F_{i\rightarrow j}$.

Notably, when $j\neq i$, $\nu_j=0$, then the potential value on $\tilde{E_j}$ will be equal to 0 also,  all of these node could be viewed as ground nodes, thus this circuit network can work well.

\subsection{Influence Matrix}\label{sec:influenceMatrix}
In light of the circuit network, we amend our linear model as follows
\begin{equation}\label{eq:linearCor}
 F_{i\rightarrow j} = \frac{1}{1+\lambda_j}\sum_{k\in N_j}{(t_{kj}F_{i\rightarrow k}+\nu_j)} \ \ \ \ for\ j=1,2,3,...n
\end{equation}
where $\nu_j$ is a correction to guarantee that, when $j=i$, $F_{i\rightarrow j}=1$. Thus, the value of $\nu_j$ could be determined as
\begin{equation}\label{eq:nu}
\nu_j=\left\{\begin{aligned}
  &a\ number\ to\ guarantee\  F_{i\rightarrow j} & \ \  &j = i& \\
  &0&\ \ \ &j\neq i&
\end{aligned}\right.
\end{equation}
Equation~\ref{eq:linearCor} could be rewritten as
\begin{equation}
  F_i=(I+\Lambda)^{-1}(T'F_i+\nuup)\nonumber
\end{equation}
where
\begin{eqnarray}
  F_i&=&[F_{i\rightarrow 1}, F_{i\rightarrow 2}, ...F_{i\rightarrow n}]'\nonumber\\
  T&=&[t_{ij}]_{n*n}\nonumber\\
  \nuup&=&[0, 0, ...\nu_i\neq 0,...0]^{T}\nonumber\\
  \Lambda&=&diag(\lambda_1, \lambda_2, ...\lambda_n)\nonumber
\end{eqnarray}
which could be solved as
\begin{eqnarray}
  \label{eq:closedsolution}
  F_i&=&(I+\Lambda-T')^{-1}\nuup\\
     &=&\Gamma^{-1}\nuup= P\nuup\label{eq:inf}
\end{eqnarray}
where the transpose of $(I+\Lambda-T')$ is strictly diagonally dominant, thus it is invertible, and we denote $(I+\Lambda-T')$ as $\Gamma=[\gamma_{ij}]_{n*n}$ and denote its inverse as $P=[p_{ij}]_{n*n}$. For $\nuup$ is a vector with only one nonzero entry $\nu_i$, thus
\begin{equation}
  F_i = \nu_i P_{\cdot i}
\end{equation}
and based on Rule 1, $F_{i\rightarrow  i}$ should be equal to 1, there is
$$
\nu_i p_{ii} = 1
$$
thus,
\begin{equation}
  \label{eq:nusolution}
  \nu_i = \frac{1}{p_{ii}}
\end{equation}
Similarly, we could get
\begin{equation}
  F_j = \nu_j P_{\cdot j}=\frac{1}{p_{jj}}P_{\cdot j}\ \ \ for\ j=1,2,3,..n
\end{equation}
This Equation could be rewritten as
\begin{eqnarray}\label{eq:influenceMatrix}
  F&=&[F_1, F_2, ...F_n]'\nonumber\\
   &=&[\frac{1}{p_{11}}P_{\cdot 1},\frac{1}{p_{22}}P_{\cdot 2},...\frac{1}{p_{nn}}P_{\cdot n}]'\nonumber\\
   &=&diag(P)^{-1}P'\nonumber\\
   &=&[f_{ij}]_{n*n}
\end{eqnarray}
Because the entries of matrix $F$ describe the influences between any pairs of nodes~(in $F$, $f_{ij}=F_{i\rightarrow j}$), in this paper we call $F$ as \textbf{influence matrix} and call $F_i$ as node $i$'s \textbf{influence vector}.

\subsubsection{The Computation of $F$.}\label{subsec:computation}
Because $F=diag(P)^{-1}P'$, the computation of $F$ is actually to get $P=(I+\Lambda-T')^{-1}$. Because $(I+\Lambda-T')$ is a strictly diagonally dominant matrix which satisfies the convergence condition of Gauss-Seidel method, its inverse $P$ could be computed by a very fast way through a Gauss-Seidel iteration process.

Because
$$
(I+\Lambda-T') P_{\cdot i}=\mathbf{e}_i,
$$
where $P_{\cdot i}$ could be viewed as the variables of this linear system of equations. For $(I+\Lambda-T')$ is strictly diagonally dominant, $P_{\cdot i}$ could be solved by Gauss-Seidel method. Specifically, Gauss-Seidel method is an iterative method which is operated as the following procedures:
\begin{enumerate}
  \item[1.] Set $p_{ji}^{(0)}=0$ for $j = 1,2,...n$;
  \item[2.] $p_{ji}^{(k+1)}=\frac{1}{\gamma_{jj}}(\mathbf{e}_{ij}-\sum_{l>j}\gamma_{jl}p_{li}^{(k)}-\sum_{l<j}\gamma_{jl}p_{li}^{(k+1)})$, for $j=1, 2, ...n$;
  \item[3.] continue Step 2 until the changes made by an iteration are below a given tolerance.
\end{enumerate}
This procedures is efficient. To get $P_{\cdot i}$ within a valid tolerance range, it often need only dozens of iterations. Thus, the time complexity of computing $P_{\cdot i}$ is $O(|E|)$ and the time complexity of $F$ is $O(|V||E|)$. Moreover, because the
computation of $P$ could be partitioned to $n$ parts, we could
compute $P$ in $n$ parts in parallel.

Even though, the computation of $F$ is still too consuming to be applied to a  large scale network. For real networks, it is not necessary to compute the influence vectors of each nodes since many of them is nobody in the network. In the following section, we will propose a fast method to estimate which nodes are important to the network.

\subsubsection{The Upper Bound of Node's Influence On a Network}\label{subsec:upperbound}
Suppose the influence vector of node $i$ is $F_i=[f_{i1},f_{i2},...f_{in}]'$, then its total influence on the network should be $\sum_{j=1}^{n}{f_{ij}}$ and we denote this number as $\mathcal{F}_{i}$ and this number could be viewed as the importance of node $i$ in the network.

On the other hand, we denote $\mathcal{P}_{i}=P_{\cdot i}\mathbf{e}=\sum_{j=1}^{n}{p_{ji}}$. Then, we could get the following property
\begin{property}\label{pr:potentialBound}
\begin{equation}\label{eq:potentialBound}
  \mathcal{F}_i\leq (1+\lambda_i)\mathcal{P}_i
\end{equation}
\begin{proof}
  For
\begin{equation}\label{eq:potential_i}
  \mathcal{F}_i = \sum_{j=1}^{n}f_{ji} =\frac{1}{p_{ii}} \sum_{j=1}^{n}p_{ji} = \frac{1}{p_{ii}}\mathcal{P}_i
\end{equation}
For $P(I+\Lambda-T')$, the dot product of
$i$-th row of $P$ and $i$-th column of $(I+\Lambda-T')$ should be 1,
that is
$$
  \sum_{l=1}^{n}{p_{il}(I+\Lambda-T')_{li}} = 1
$$
\noindent where $(I+\Lambda-T')_{ii}=(1+\lambda_i)$ and $(I+\Lambda-T')_{li}=-t_{li}$ for any $l\not{=}i$. Then, this equation is equivalent to $
p_{ii}(1+\lambda_i)-\sum_{l=1,l\not{=}i}^{n}{p_{il}t_{li}}=1 $ .
Thus, we have $
p_{ii}(1+\lambda_i)=1+\sum_{l=1, \,l\not{=}i}^{n}{p_{il}t_{li}}\geq 1
$. And then,
$$
  \frac{1}{p_{ii}}\leq (1+\lambda_i)
$$
With Equation~\ref{eq:potential_i}, there is
$$
  \mathcal{F}_i \leq (1+\lambda_i)\mathcal{P}_i
$$
\end{proof}
\end{property}
In Equation~\ref{eq:potentialBound}, the quantity $\mathcal{P}_i$ is a value that could be got by fast method. Let's denote $\mathcal{P}=[\mathcal{P}_1,\mathcal{P}_2,...\mathcal{P}_n]'$, for $\mathcal{P}_i = P_{\cdot i}\mathbf{e}$, there is $\mathcal{P}=P'\mathbf{e}$ which could be rewritten as
$$
(I+\Lambda-T)\mathcal{P}=\mathbf{e}
$$
This is a linear system of equations which satisfy the convergence condition of Gauss-Seidel method. As the variables of this system, $\mathcal{P}$ could be computed in $O(|E|)$ time by the similar way described in Section~\ref{subsec:computation}. Thus, thanks to Property~\ref{pr:potentialBound}, we could get the upper bound of nodes' importance in $O(|E|)$ time. These upper bounds could help us to quickly identify those less-importance nodes and skip them. What's more, $\mathcal{F}_i$ is a very compact upper bound to estimate the importance of node $i$ and we will demonstrates it in Section~\ref{sec:exp}.

\section{The Influence From A Node Set}\label{sec:catTwo}
In real applications, we may need to compute the influence from a node set $S$ to the nodes of the network~(Consistent with the notation
$F_{i\rightarrow j}$, we denote the influence from set $S$ to node $j$ as $F_{S\rightarrow j}$). In~\cite{goyal2010learning}, Goyal et al assumed that the influences of different nodes are
independent of each other. Hence, the joint influence $F_{S\rightarrow j}$ can
be defined as $$
  F_{S\rightarrow j} = 1 - \prod_{k\in S}(1-F_{k\rightarrow j})
$$
However, the influences from different nodes are obviously not mutually
independent. Then, how can we get the independent influence of a node when it belongs to a node set $S$?

\subsection{Independent Influence Inference}\label{sec:indInfInf}

To simplify the discussion, we will first turn to solve an
equivalent problem defined as follows.

\begin{problem}\nonumber\label{Pr:IndInf}
Given a seed set $S$ and a node $k\not{\in} S$, how to evaluate the independent influence $F_{k\rightarrow j}^{(S)}$ (independent from the nodes of $S$)?
\end{problem}

For this problem, if we could work out $F_{k\rightarrow j}^{(S)}$, then we could work out $F_{s\rightarrow j}^{(S+\{k\}-\{s\})}$ for any $s\in S$ in the same way. Then, for the event source set $S'=S+\{k\}$, we could work out
the independent influence for any element in $S'$, and
\begin{equation}\label{eq:seedsetIF}
  F_{S'\rightarrow j}=1 - \prod_{i\in S'}(1-F_{i\rightarrow j}^{(S'-\{i\})})
\end{equation}
Because $F_{i\rightarrow j}= f_{ij}$, for ease of writing and consistent with the previous notation, in the following text, we will use $f_{ij}$ to denote
$F_{i\rightarrow j}$, use $f_{S'j}$ to denote $F_{S'\rightarrow j}$ and use $f_{ij}^{(S)}$ to denote $F_{i\rightarrow j}^{(S)}$. By Equation~\ref{eq:seedsetIF}, the happening
probability $f_{S'j}$ can be computed. Thus, Problem \ref{Pr:IndInf} is
the basis for computing the influence from a node set. In order
to get the independent influence $f_{kj}^{(S)}$, let us think about
the essence of independent influence in light of the circuit perspective.
We could summarize the independent influence into two conditions:
\begin{enumerate}
  \item Each node in $S$ will never be influenced by node $k$;
  \item Each node in $S$ will never propagate the influence derived from node $k$.
\end{enumerate}

Because the nodes in $S$ are the ones on which event $e$ has already
happened, they do not need to be influenced by $k$ and cannot be
influenced. Therefore, the above condition 1 needs to be satisfied.
On the other hand, because each node in $S$ itself will spread the
influence on the network about event $e$, it will block the same
type of influence from $k$. Therefore,  we assume nodes in $S$ will
never propagate the influence from $k$.

\begin{figure}[!h]
\begin{center}\includegraphics[width=0.28\textwidth]{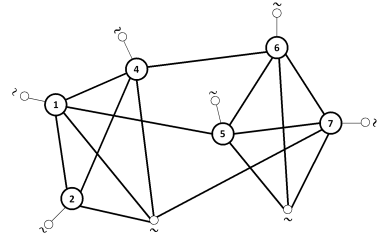}
\caption{The circuit network corresponding to Figure
1 (a).} \label{fig:minor-modified circuit network}
\end{center}
\end{figure}

Based on the above two conditions, we could construct a circuit
network $G'(S,k)$ to model the independent influence from node $k$
to the other nodes as follows:
\begin{enumerate}
  \item First, construct the circuit network $G'$;
  \item Put an electric pole with voltage value 0 on each node in $S$;
  \item Set the voltage value on external electrode $\tilde{E_k}$ to $\frac{\nu^{(S)}_i}{(1+\lambda_i-\theta_i)}$;
  \item Set the voltage value on external electrode $\tilde{E_i}(i\neq k)$ to 0;
\end{enumerate}
For the graph in Figure~\ref{fig-network}(a), suppose the seed set
$S=\{3,8\}$ and $k=4$, then the corresponding $G'(S,4)$ is showed in
Figure~\ref{fig:minor-modified circuit network}. From this figure,
we can see that the above two conditions about independent influence
are satisfied naturally.

\subsection{The Deduction of Independent Influence}\label{sec:indinfDec}

In the previous subsection, we have constructed the circuit network
for simulating the independent influence. In this subsection, we first analyze the potential value on each node which could be
viewed as the influence on it. Then, we provide
several properties of independent influence and the efficient way
for computing it.

\textbf{The Electric Potential on Each Node for $G'(S,k)$}.
Based on the similar discussion as shown in
subsection~\ref{sec:influenceMatrix}, we know that the electric
potential on node $j$ (except for $S$) is
\begin{equation}\label{eq:kiffCat2}
  \tilde{U}_j
=\frac{1}{1+\lambda_j}(\sum_{l=1}^{n}{t_{lj}\tilde{U}_{l}}+\nu_j^{(S)})\ \ \ \ \ \ \mathrm{for}\ j\not{\in{S}}.
\end{equation}
where
\begin{equation}
  \label{eq:nuS}
  \nu_j^{(S)}=
                          \left\{\begin{aligned}
  &a\ number\ to\ make\ \tilde{U}_j=1&\ \ \ \ \  &j=k& \\
  &0&\ \ \ \ \ \ \ \ &otherwise&
\end{aligned}\right.
\end{equation}
Because
$
\tilde{U}_l=0,  \mathrm{for}\ l\in S
$, then Equation~\ref{eq:kiffCat2} can be rewritten as
\begin{equation}\label{eq:kiffCat2.2}
(1+\lambda_j)\tilde{U}_j-\sum_{l\not{\in}S}t_{jl}\tilde{U}_l= \nu_j^{(S)} \ \ \ for\ j\not{\in{S}}.
\end{equation}

Without loss of generality, we can set $S=\{n-|S|+1, n-|S|+2,...n\}$
(i.e., the last $|S|$ nodes in $G'$), then Equation~\ref{eq:kiffCat2.2}
can be rewritten as
$
(1+\lambda_j)\tilde{U}_j-\sum_{1}^{n-|S|}t_{jl}\tilde{U}_l = \nu_j^{(S)}, \ \mathrm{for}\ j=1,2,...,n-|S|
$, which is equivalent to

\begin{equation}\label{eq:indUarray}
  \Gamma_{\overline{S}\overline{S}}\tilde{U}_{\overline{S}} = \nuup_{\overline{S}}
\end{equation}

\noindent where $\tilde{U}_{\overline{S}}=[\tilde{U}_{1},
\tilde{U}_{2},...\tilde{U}_{n-|S|}]^T$,
$\nuup_{\overline{S}}=[\nu_{1}, \nu_{2},...\nu_{n-|S|}]^T$ and
$\Gamma_{\overline{S}\overline{S}}$ is a matrix which is cut
down from $\Gamma$ by removing the columns and rows from
$n-|S|+1$ to $n$.

$$
\Gamma_{\overline{S}\overline{S}}=\left[\begin{aligned}
  &\gamma_{11}&\ &\gamma_{12}&\ ...&\gamma_{1(n-|S|)}&\\
&...&\ &...& \ &...&\\
&\gamma_{(n-|S|)1}&\ &\gamma_{(n-|S|)2}&\ &\gamma_{(n-|S|)(n-|S|)}&
\end{aligned}
\right ].
$$

and then, by Equation~\ref{eq:indUarray}, we have

\begin{equation}\label{eq:indU}
\tilde{U}_{\overline{S}}=\Gamma_{\overline{S}\overline{S}}^{-1}\nuup_{\overline{S}}
\end{equation}

For only the $k$-th element of $\nuup^{(S)}$ is nonzero, $\tilde{U}_{\overline{S}}$ is equal to the
multiplication of $\nu_k^{(S)}$ and the $k$-th column of
${\Gamma^{-1}_{\overline{S}\overline{S}}}$.

Based on Equation~\ref{eq:nuS}, we know that the number of $\nu_k^{(S)}$ should guarantee that $\tilde{U}_k=1$. By
Equation~\ref{eq:indU}, that is
$(U_{\overline{S}})_k=\nu_k^{(S)}(\Gamma_{\overline{S}\overline{S}}^{-1})_{kk}=1$.
Then,
\begin{equation}
\nu_k^{(S)}=\frac{1}{(\Gamma_{\overline{S}\overline{S}}^{-1})_{kk}}
\end{equation}
And we could get
\begin{equation}\label{eq:indInf}
    (U_{\overline{S}})_j=\left\{\begin{aligned}
    &\frac{(\Gamma_{\overline{S}\overline{S}}^{-1})_{jk}}{(\Gamma_{\overline{S}\overline{S}}^{-1})_{kk}}&\ \ \ \ &j=1,2,...n-|S|&\\
      &0&\ \ \ \ &otherwise&
    \end{aligned}\right .
  \end{equation}
Based on the previous discussion, the potential on node $j$ could be viewed as the independent influence from node $k$ to node $j\not\in S$. Thus, there is
\begin{theorem}\label{th:indInf}
  \begin{equation}\label{eq:indInf}
    f_{kj}^{(S)}=\left\{\begin{aligned}
    &\frac{(\Gamma_{\overline{S}\overline{S}}^{-1})_{jk}}{(\Gamma_{\overline{S}\overline{S}}^{-1})_{kk}}&\ \ \ \ &j=1,2,...n-|S|&\\
      &0&\ \ \ \ &otherwise&
    \end{aligned}\right .
  \end{equation}
\end{theorem}
Theorem~\ref{th:indInf} for $k=1,2,...n-|S|$ can be rewritten as
$$
  F^{(S)}=diag(\Gamma_{\overline{S}\overline{S}}^{-1})(\Gamma_{\overline{S}\overline{S}}^{-1})'
$$.
\textbf{The Properties of Independent Influence.} We have two properties about the independent influence as follows.
\begin{property}\label{pr:indInfG}
$
\ \ \ \ \ \   f_{kj}^{(S)}\leq f_{kj}.
$
\end{property}
\begin{property}\label{pr:indPotential}
 If define $\mathcal{F}_k^{(S)}=\sum_{i=1}^{n-|S|+1}{f_{kj}^{(S)}}$, then
 $\ \mathcal{F}_k^{(S)}<\mathcal{F}_k<(1+\lambda_k)\mathcal{P}_k$.
\end{property}

Property~\ref{pr:indInfG} shows that $f_{kj}^{(S)}$ cannot be
greater than $f_{kj}$. If $S$ is not empty, the influence from
node $k$ will be diluted by the influence from nodes in $S$.
Property~\ref{pr:indPotential} provides the upper bound of
$\mathcal{F}_k^{(S)}$ which can help us reduce the unnecessary
computation for those nodes with small independent influence in
the network.

\textbf{The Computation of Independent Influence.} Based on
Theorem~\ref{th:indInf}, given the seed set $S$, if we want to
compute the independent influence of node $k$, we just need to
compute the $k$-th column of
$\Gamma_{\overline{S}\overline{S}}^{-1}$. Because $\Gamma_{\overline{S}\overline{S}}$ is a strictly diagonally dominant matrix, it satisfies the convergence condition of
Gauss-Seidel method. And from
$\Gamma_{\overline{S}\overline{S}}{\Gamma_{\overline{S}\overline{S}}^{-1}}=I$, there is
$\Gamma_{\overline{S}\overline{S}}{\Gamma_{\overline{S}\overline{S}}^{-1}}_{\cdot
k}=\mathbf{e}_k$. Taking the similar procedures in Section~\ref{subsec:computation}, we could compute the $k$-th column of
${\mathcal{L}_{\overline{S}\overline{S}}^{-1}}$ in $O(|E|)$ time.

\section{Social Influence Maximization}\label{sec:problem}

In this section, we show how to exploit the circuit model for handling the
social influence maximization problem, which targets on finding a
set of seeds in a way that these seeds will influence the maximal
number of individuals in the network.

\subsection{Problem Formulation and Algorithm}
The social influence maximization problem can be formalized
as
\begin{equation}
  S={\arg\max}_{S\subset V}\sum_{j=1}^{n}F_{S\rightarrow j}\ \ \ \ subject\ to\ \ |S|=K
  \nonumber
\end{equation}
where $F_{S\rightarrow j}$ follows the meaning in Equation~\ref{eq:seedsetIF} which is the influence from $S$ to node $j$. And in the following text, we denote $(\sum_{j=1}^{n}F_{S\rightarrow j})$ as $\sigma(S)$ which is the influence spread function of $S$(i.e. the number of individuals will be influenced by $S$).

\textbf{Properties of $\sigma(S)$.} Under the circuit model, we
could prove that the influence spread function $\sigma(S)$ is a
submodular function. This shows in the following theorem.
\begin{theorem}\label{th:submodular}
  For all the seed sets, $S_1\subseteq S_2\subseteq V$ and node $s$, it holds that
  \begin{equation}\label{eq:submodular}
    \sigma(S_1\cup \{s\})-\sigma(S_1)\geq \sigma(S_2\cup \{s\})-\sigma(S_2)
  \end{equation}
  \begin{proof}
    For $\sigma(S)=\sum_{j=1}^{n}F_{S\rightarrow j}={\sum}_{j=1}^{n}f_{Sj}$, and then,
    \begin{eqnarray}
      \sigma(S\cup \{s\})&=&{\sum}_{j=1}^{n}(1-(1-f_{Sj})(1-f_{sj}^{(S)})) \nonumber \\
&=&{\sum}_{j=1}^{n}(f_{sj}^{(S)}+f_{Sj}-f_{sj}^{(S)}f_{Sj}))\nonumber
    \end{eqnarray}
thus
$$
\sigma(S\cup
\{s\})-\sigma(S)={\sum}_{j=1}^{n}(f_{sj}^{(S)}(1-f_{Sj}))
$$
For $S_1\subseteq S_2$, it is easy to get that
$$
f_{sj}^{(S_1)}\geq f_{sj}^{(S_2)},\ \  f_{S_1j}\leq f_{S_2j}
$$
and then,
$$
{\sum}_{j=1}^{n}(f_{sj}^{(S_1)}(1-f_{S_1j}))\geq {\sum}_{j=1}^{n}(f_{sj}^{(S_2)}(1-f_{S_2j}))
$$
which is,
$$\sigma(S_1\cup \{s\})-\sigma(S_1)\geq \sigma(S_2\cup
\{s\})-\sigma(S_2)$$.
  \end{proof}
\end{theorem}

For simplicity, we denote $\sigma(S\cup
\{s\})-\sigma(S)=\mathfrak{F}_s^{(S)}$ in the following text. From the proof of
Theorem~\ref{th:submodular}, $
\mathfrak{F}_s^{(S)}={\sum}_{j=1}^{n}(f_{sj}^{(S)}(1-f_{Sj})) $, specifically,
$\mathfrak{F}_s^{(\emptyset)}=\mathcal{P}_s$. Thus, Equation~\ref{eq:submodular} could be rewritten as $\mathfrak{F}_s^{(S_1)}\geq \mathfrak{F}_s^{(S_2)}$.

\begin{algorithm} [th]
\SetKwInOut{Input}{input}
\SetKwInOut{Output}{output}
\Input{$G(V,A,C),K, \lambda$\\}
\Output{$S$\\}
Compute $\mathcal{P}=[\mathcal{P}_1,\mathcal{P}_2,...\mathcal{P}_n]'=(I+\Lambda-T)^{-1}\mathbf{e}$~(see it in Section~\ref{subsec:upperbound});\\
\For{each vertex $i$ in $G$}
{
    $\mathfrak{F}_i=(1+\lambda_i)\mathcal{P}_i$;
}
\While{$|S|<K$}
{
    re-arrange the order of vertex to make $\mathfrak{F}_i\geq \mathfrak{F}_{i+1}$;\\
    $\mathfrak{F}_{max}=0$;\\
    \For{int $i=1$ to $n-|S|$}
    {
        \If{$\mathfrak{F}_i>\mathfrak{F}_{max}$}
        {
            //update $\mathfrak{F}_i^{(S)}$  \\
            $\mathfrak{F}_i = UpdateIncrement(i,S,\Lambda,T, G,F_S)$; \\
            \If{$\mathfrak{F}_i>\mathfrak{F}_{max}$}
            {
                $\mathfrak{F}_{max}=\mathfrak{F}_i$;\\
                $s=i$;\\
            }
        }
        \Else
        {
            break;\\
        }
    }
    \For{int $i=1$ to $n-|S|$}
    {
        $f_{Si}=1-(1-f_{Si})(1-f_{si}^{(S)})$//update $F_S$\\
    }
    $S\leftarrow S\cup \{s\}$;\\
    $\mathfrak{F}_s \leftarrow 0$;\\
}
return $S$;\\
\caption{Circuit($G$, $K$, $\Lambda$, $T$)}\label{algo:Circuit}
\end{algorithm}

\textbf{Proposed Algorithm.} By a greedy strategy framework, we
always choose the node which can produce the most increment on
influence spread when adding it into $S$. The greedy algorithm
starts with an empty set $S_0=\emptyset$, and iteratively, in step
$k$, adds the node $s_k$ which maximizes the increment on influence
spread into $S_{k-1}$
$$
s_k = {\arg\max}_{s\in V\setminus S_{k-1}}\mathfrak{F}_s^{(S_{k-1})}
$$
In this process, there is
\begin{property}\label{pr:potentialseqGeq}
$
\ \ \ \mathcal{P}^{max}_s\geq\mathfrak{F}_s^{(S_0)}\geq\mathfrak{F}_s^{(S_1)}\geq\mathfrak{F}_s^{(S_2)}...\geq\mathfrak{F}_s^{(S_K)}.
$
\end{property}

Based on these properties, we design an algorithm to solve the
social influence maximization problem, as shown in
Algorithm~\ref{algo:Circuit}. In this algorithm, we use
$\mathfrak{F}_i$ to store the upper bound of the node $i$'s
authority which can be used to indicate whether or not the node is
an important node, and use $\mathfrak{F}_{max}$ to store the
current maximum increment. For example, in step $k$, suppose the
current maximum increment is $\mathfrak{F}_{max}$, and when we run
into a node $j$, if its increment in last step
$\mathfrak{F}_j^{(S_{k-1})}\leq\mathfrak{F}_{max}$, then we do not
need to compute its current increment
$\mathfrak{F}_j^{(S_{k})}$ since it is impossible for
$\mathfrak{F}_j^{(S_{k})}$ to be larger than its previous increment
$\mathfrak{F}_j^{(S_{k-1})}$; otherwise, we first update
$\mathfrak{F}_i^{(S_{k-1})}$ to $\mathfrak{F}_i^{(S_{k})}$, if
$\mathfrak{F}_i^{(S_{k})}$ is greater than $\mathfrak{F}_{max}$, we
assign $i$ to a store variable $s$ to keep it and assign
$\mathfrak{F}_j^{(S_{k})}$ to $\mathfrak{F}_{max}$; at last, the
value in $s$ will be the expected $s_k$ and we add it into the seed
set $S$. In this process, we keep a subprocedure to compute
$F_S=[f_{S1},f_{S2},...f_{Sn}]$ which will be used in the update of
$\mathfrak{F}_i^{(S_{k})}$ in Function~\ref{func:updateAuthority}.
In addition, in every step (i.e., for finding a seed), we
need to re-arrange the order of node to make $\mathfrak{F}_{i}\geq
\mathfrak{F}_{i+1}$ which can also help to further reduce the
computational cost. Function~\ref{func:updateAuthority} is used to compute the influence/independent-influence of node and return the $(i,S)$-Authority of node $i$.

\begin{function}
\SetKwInOut{Input}{input} \SetKwInOut{Output}{output}
\Input{$i$, $S$, $\Lambda$, T, $G$, $F_S$}
\Output{$\mathfrak{F}_i$\\}
Compute the $i$-th column of $\Gamma_{\overline{S}\overline{S}}^{-1}$ by the methods in Section~\ref{sec:indinfDec} and then get $f_{ij}^{(S)}$ based on Equation~\ref{eq:indInf}.\\
\For{int $j=1$ to $n-|S|$}
{
    //Based on the discussion in Theorem~\ref{th:submodular},\\
    $\mathfrak{F}_i += f_{ij}^{(S)}(1-f_{Sj})$;\\
}
return $\mathfrak{F}_i$;\\
 \caption{UpdateIncrement($i$, $S$, $\Lambda$, T, $G$, $F_S$)}\label{func:updateAuthority}
\end{function}

\begin{figure*}
  \begin{center}
    \subfigure[ \vspace{-1cm}  Wiki-Vote.
    \quad]{\label{fig:wiki}\includegraphics[scale=0.18]{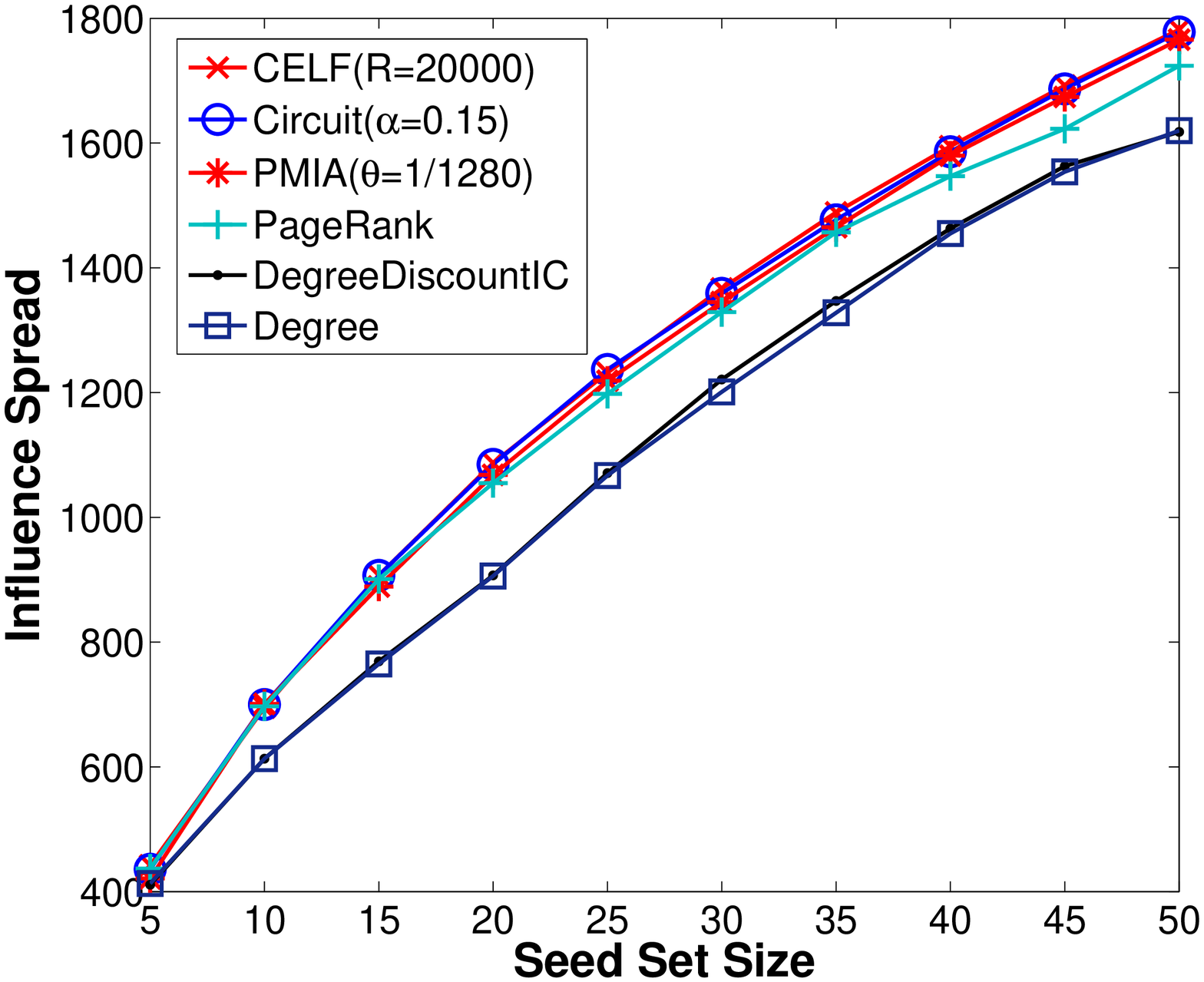}}  
    \subfigure[ \vspace{-1cm}  ca-HepPh.
    \quad]{\label{fig:phy}\includegraphics[scale=0.18]{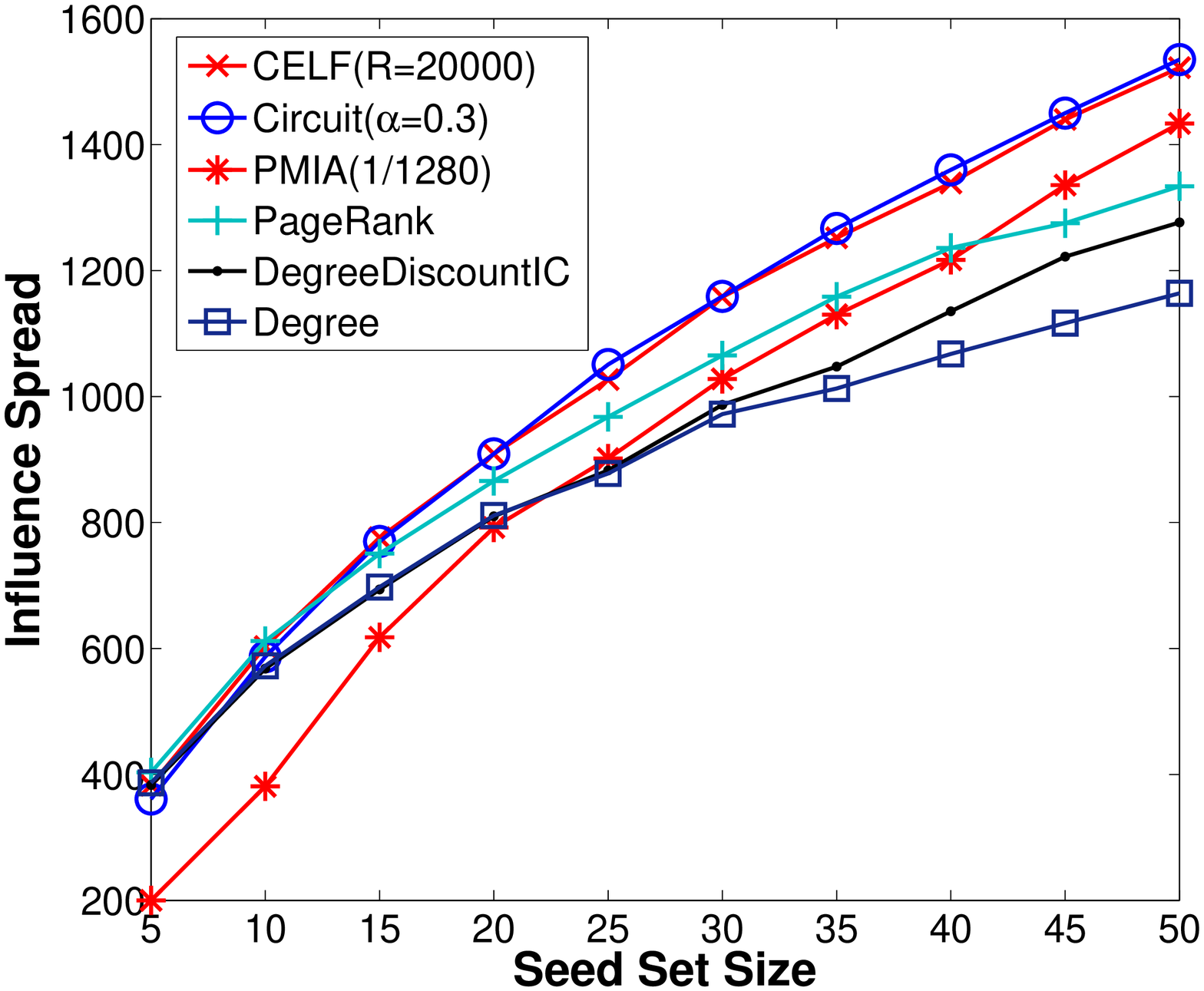}}   
    \subfigure[ \vspace{-1cm}  DBLP.
    \quad]{\label{fig:dblp}\includegraphics[scale=0.18]{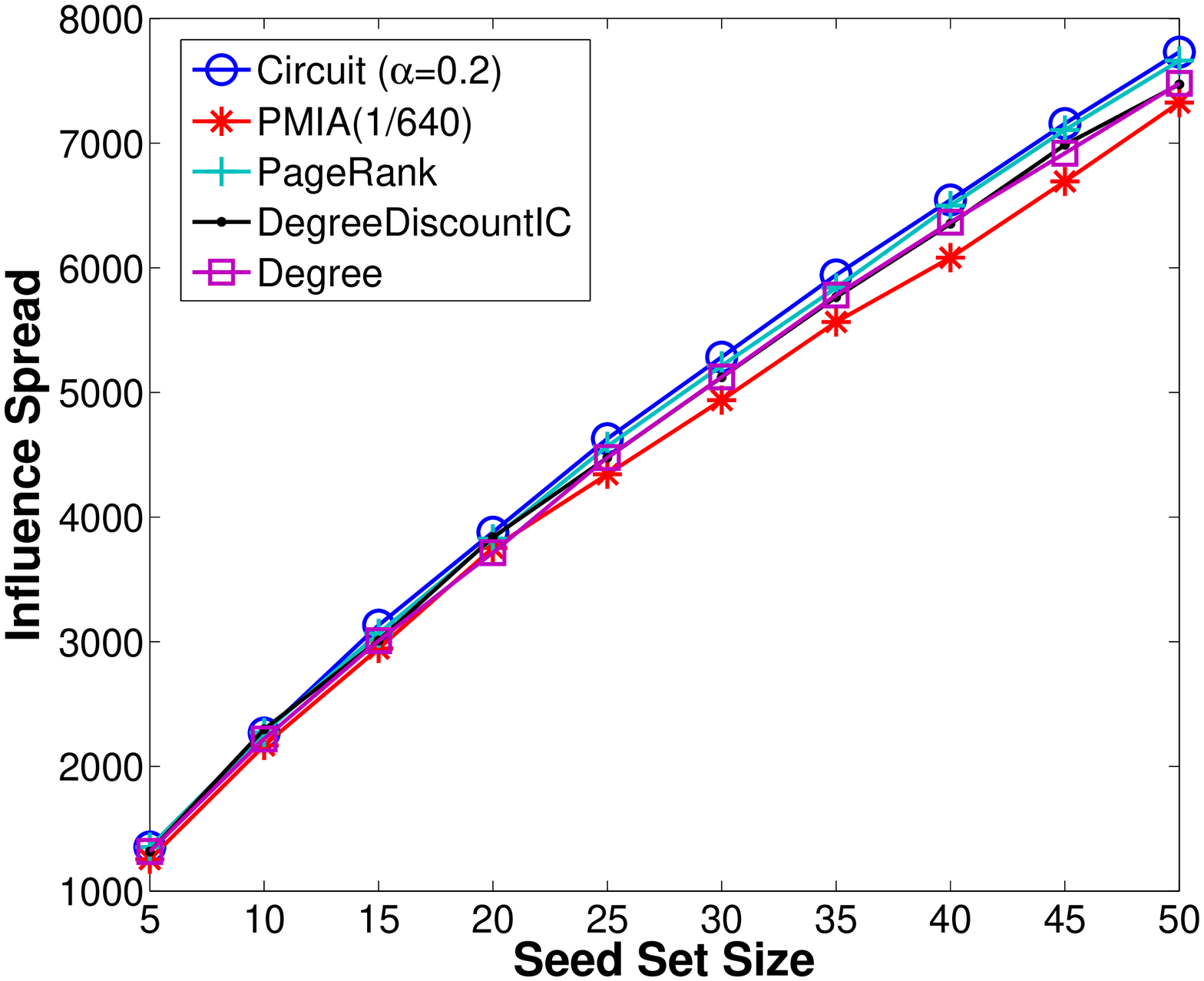}}   
    \subfigure[ \vspace{-1cm}  Amazon.
    \quad]{\label{fig:amazon}\includegraphics[scale=0.18]{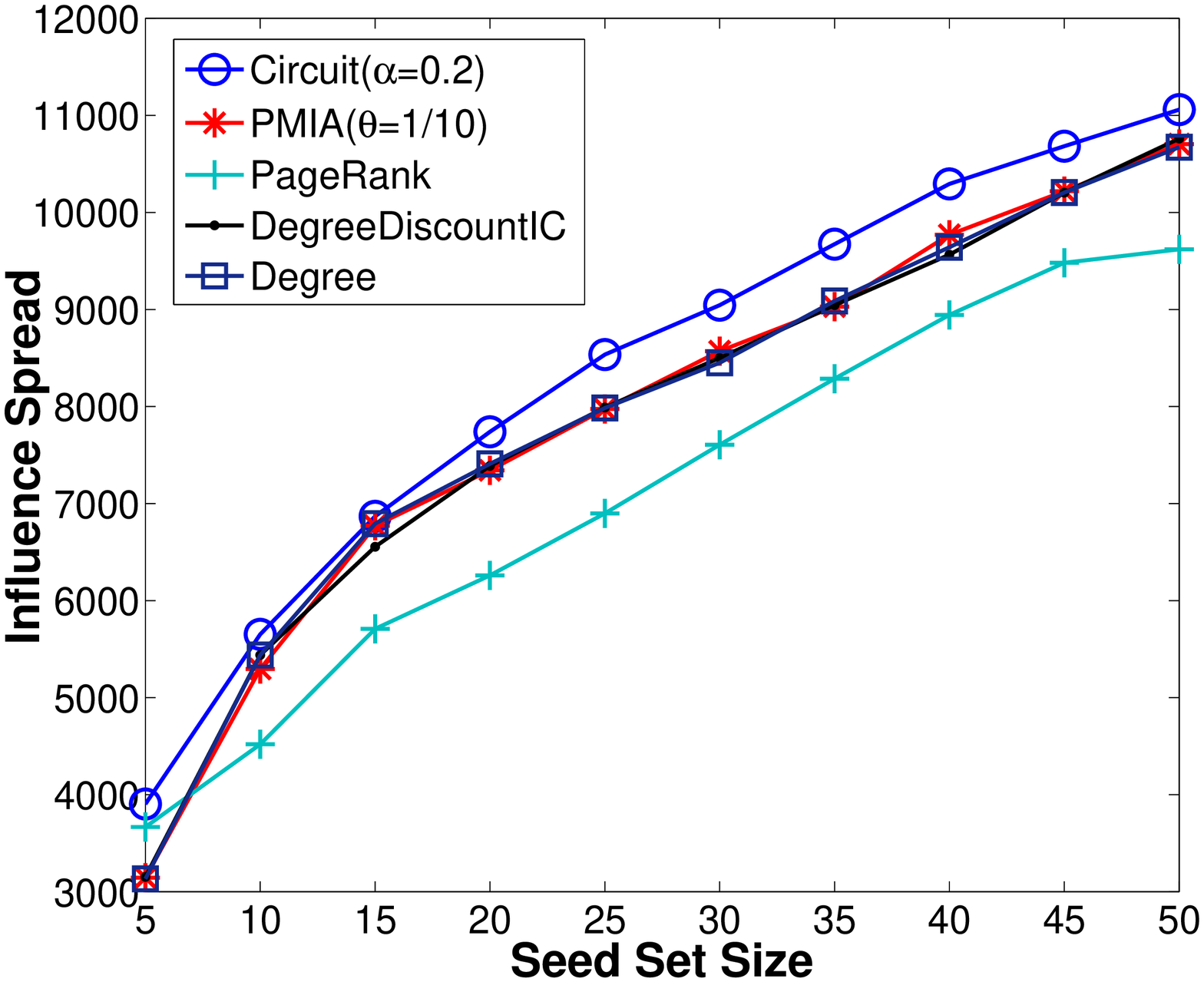}}  
  \end{center}
  \caption{The results of influence spread on five benchmark network datasets. \label{fig:exp}}
\end{figure*}

\section{Experimental Results}\label{sec:exp}
Here, we evaluate the performances of the  Circuit method on
several real-world social networks. Specifically, we demonstrate:
(a) the scalability and the ability to maximize the influence propagation comparing to benchmark
algorithms; (b) the impact of the damping coefficient $\lambda$; and
(c) the effectiveness of the upper bounds.

\subsection{The Experimental Setup}

\textbf{Experimental Data.} Four real-world networks have been used in the experiments. The
first one is a Wikipedia voting network in which nodes represent
wikipedia users and a directed edge from node $i$ to node $j$
represents that user $i$ voted on user $j$, the network contains all
the Wikipedia voting data from the inception of Wikipedia till
January 2008~\footnote{http://snap.stanford.edu/data/wiki-Vote.html}.
The second one, denoted as
ca-HepPh~\footnote{http://snap.stanford.edu/data/ca-HepPh.html}, is a collaboration network which is from the e-print arXiv, and
it covers scientific collaborations between authors whose papers
have been submitted to \emph{High Energy Physics - Phenomenology category}.  The third one is an even larger collaboration
network, the DBLP Computer Science Bibliography Database, which is the same as in~\cite{chen2010scalable}. The
fourth one is another large directed network collected by crawling
the Amazon website. It is based on \emph{Customers Who Bought This Item
Also Bought} feature of the Amazon website , where if a product $i$
is frequently co-purchased with product $j$, then a directed edge
from $i$ to $j$ will be
added~\footnote{http://snap.stanford.edu/data/amazon0302.html}. We
chose these networks since they can cover a variety of networks
with sizes ranging from 103K edges to 2M edges and include two
directed networks and two undirected networks. Some basic
statistics about these networks are shown in
Table~\ref{table:datasets}.

\begin{table}[bhpt] \centering \caption{Statistics of five selected
real-world networks.} \scriptsize\label{table:datasets}
\begin{center}
\begin{tabular}{|c|l|l|l|l|}
 \hline
  Networks  &  Wiki-Vote & ca-HepPh   &  DBLP  & Amazon \\
\hline\hline \#Node & 7,115 & 12,008 & 655K & 262K\\ \hline

\#Edge/Arc & 103,689 & 237,010 & 2.0M & 1.2M\\ \hline

Type & directed & undirected & undirected & directed\\ \hline
\end{tabular}
\end{center}
\end{table}

\begin{figure}
  \begin{center}\hspace{-1cm}
  \includegraphics[scale=0.30]{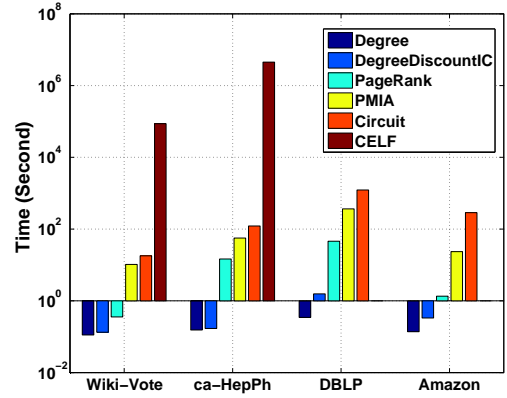} \vspace{-0.4cm}\caption{The computational performances. \label{fig:time}}
  \end{center}
\end{figure}

\textbf{Benchmark Algorithms.} The benchmark algorithms are as follows. First, \textbf{Circuit} is our Algorithm~\ref{algo:Circuit}, where each node's influence/ independent-influence will be computed as the average result of 10 iterations and we set damping coefficient matrix $\Lambda=\lambda I$ and $\lambda$ ranges in $(0,1)$~\footnote{The number of $\lambda$ should be larger than 0 and smaller than 1, and the reason is omitted for the limited space.}. {\textbf{CELF}(20000)} is the original greedy algorithm with the CELF optimization of \cite{leskovec2007cost}, where  R = 20000. \textbf{PMIA($\theta$)} is the algorithm proposed in~\cite{chen2010scalable}. We used the source code provided by the
authors, and set the parameters to the ones produces the best results~\footnote{Based on the source code from its author, the parameter would be selected from \{1/10,1/20,1/40,1/80,1/160,1/320,1/1280\}}. In the \textbf{PageRank} algorithm~\cite{page1999pagerank}, we selected top-$K$ nodes with the highest pagerank value. \textbf{DegreeDiscountIC}~\cite{chen2009efficient} measures the degree discount heuristic  with a propagation probability of $p = 0.01$, which is the same as used in \cite{chen2009efficient}. Finally, the \textbf{Degree} method captures the top-$K$ nodes with the highest degree. Among these algorithms, Degree, DegreeDiscountIC and pageRank are
widely used for baselines. To the best of our knowledge, CELF and
PMIA are two of the best existing algorithms in terms of solving
the influence maximization problem~(concerning the tradeoff between
effectiveness and efficiency).

\textbf{Measurement.} The effectiveness of the algorithms for the social influence maximization problem is justified by the number of nodes that will be activated by the seed set of the algorithm. This is called \textbf{influence spread}. To obtain the influence spread of these algorithms, for each seed
set, we run the Monte-Carlo simulation under the Weighted
Cascade (WC) model~\footnote{Other models are also introduced in ~\cite{kempe2003maximizing}, but due to the space limitation, we focus on the WC model, which is an important case of the Independent Cascade(IC) model. What's more, we can prove that the amounts of the IC models could be reformed as the WC model. The proof is also included in our full technical report.}  20000 times to find how many nodes can be
influenced~\footnote{In detail, under the WC model, the node in the seed set propagates its influence through the following operations. Let us view the node in the seed set $S$ as the node activated at time $t=0$,  if node $i$ is activated at time $t$, then it will activate its not-yet-activated neighbor node $j$ at time $t+1$ (and only time $t+1$) with the probability $\frac{c_{ji}}{d_j}$.}, and then use these influence spreads to
compare the effectiveness of these algorithms. Since the CELF algorithm is very time-consuming, and DBLP and
Amazon networks are too large for it to handle. Thus, we just get
 the experimental results for CELF on three comparatively smaller networks.

{\bf Experimental Platform}. The experiments were performed
on a server with 2.0GHz Quad-Core Intel Xeon E5410 and 8G memory.

\subsection{A performance comparison}
In the following, we present a performance comparison of both
effectiveness and efficiency between Circuit and the benchmarks. For
the purpose of comparison, we record the best performance of each
algorithm by tuning their parameters. We run tests on the five
networks under the WC model to obtain the results of influence spread. The seed
set size $K$ ranges from 1 to 50. Figure~\ref{fig:exp} shows the
final results of influence spread. For easy to read, we paint tokens at
each 5 points. Figure~\ref{fig:time} shows the computational performance
comparison for selecting 50 seeds.

In Figure~\ref{fig:wiki}, we can observe that CELF, Circuit and
PMIA are three best algorithms for Wiki-Vote. PageRank performs well
at the beginning but deteriorate when the seed set
size is large~(e.g., larger than 35).  DegreeDiscountIC and
Degree perform worse than others and
DegreeDiscountIC performs a little bit better than Degree. Similar results can also be observed from the rest networks.

In summary, in most cases, Circuit and CELF
perform much better than other methods. We believe there are two
reasons. First, for these algorithms, their focus on dealing
with the independent influence are very different. Degree and
PageRank do not care whether the node's influence is
independent from other nodes or not. While DegreeDiscountIC and PMIA try
to remove the influence that is diluted by other nodes, their
methods are too simple to produce the real independent influence.
In contrast, by a reasonable and tractable method, Circuit is able
to compute each node's independent influence, and please note that
CELF may also get each node's real independent influence if we set
the parameter R to an unlimited number. Thus, we can say that
the more attention the algorithm pays on the independent influence,
the better results the algorithm may get. Second, the structure
of the networks are quite different (e.g. clustering coefficient),
and these differences may also lead to different performances for
each algorithm. For example, we find that Degree performs
well for the networks where the cluster coefficient are comparably
small~(in such networks, the most influential nodes often
belong to different clusters, and their influence will almost
independent from each other, so Degree can
work well).

Figure~\ref{fig:time} shows the running time comparison for
selecting 50 seeds on the five networks. We can observe that the
order of the algorithms in running time is Degree $>$ DegreeDiscountIC
$>$ PageRank
$>$ PMIA $>$ Circuit $>$ CELF, where ``$>$'' means ``is more efficient than''. We can see that CELF is not scalable to large networks with million
edges; while Circuit is scalable, and it takes less than one hour to
run on the DBLP network with 2.0M edges. By comparing Figure~\ref{fig:exp}
and Figure~\ref{fig:time}, we can observe that there are some kind of
correlations between the effectiveness and efficiency of each
algorithm. Thus, the differences in running time of these algorithms
should also mainly lie in their solutions to independent influence.
Generally speaking, the more attention they pay on the independent
influence, the more complex the algorithm, and more time will take for running the results.

\begin{table}[bhpt]\centering \caption{Detailed Comparison Between
Circuit and CELF.} \scriptsize\label{table:characteristics}
\begin{center}
  \begin{tabular}{|c|l|l|l|l|}
 \hline
Network    &  Effectiveness   &  Speedup   &  Search Ratio \\
\hline\hline Wiki-Vote & 0.996 & 1578 & 0.037 \\ \hline
ca-HepPh & 1.006 & 4303 & 0.035 \\
\hline
\end{tabular}
\end{center}
\end{table}

\textbf{A Comparison with CELF.} Here, we provide a more
detailed comparison between Circuit and CELF, the best two
algorithms in term of effectiveness. Specifically, we compare their
average performances on the three networks, and the results are shown
in Table~\ref{table:characteristics}, where the performance of CELF
is chosen as the baseline. In this table, ``Effectiveness'' is the
ratio of Circuit's influence spread result relative to CELF's,
``Speedup'' is the ratio of Circuit's running time relative to
CELF's and ``Search Ratio'' is the ratio of the number of nodes
which have been investigated by Circuit relative to the number of
nodes investigated by CELF. Similar to Figure~\ref{fig:exp}, we
can see that the Circuit's effectiveness is very close to CELF in
three moderate networks: 0.4\% worse than CELF in Wiki-Vote but
0.6\% better than CELF in ca-HepPh.
For efficiency, Circuit outperforms CELF significantly, in these three
networks, it is 1578, 835, and 4303 times faster than CELF
respectively. With respect to the number of search nodes, CELF needs
to search all nodes in the network, while Circuit just need to
search less than 5\% of nodes due to the effect of
Property~\ref{pr:potentialBound}.

\textbf{Summary.} Generally, for solving the social influence
maximization problem, Circuit and CELF perform consistently well on
each network, but CELF is not scalable to large networks. Though
PMIA and PageRank are two good scalable heuristic algorithms, their
performances are not stable for many networks~(e.g., DBLP and Amazon
in this paper), and similar results can be also observed for both
Degree and DegreeDiscountIC.

\subsection{The Impact of $\lambda$}

\begin{figure}
  \begin{center}
    \vspace{0.3cm}
    \subfigure[ \vspace{-1cm}  Wiki-Vote.
    \quad]{\label{fig:wikiEffi}\includegraphics[scale=0.18]{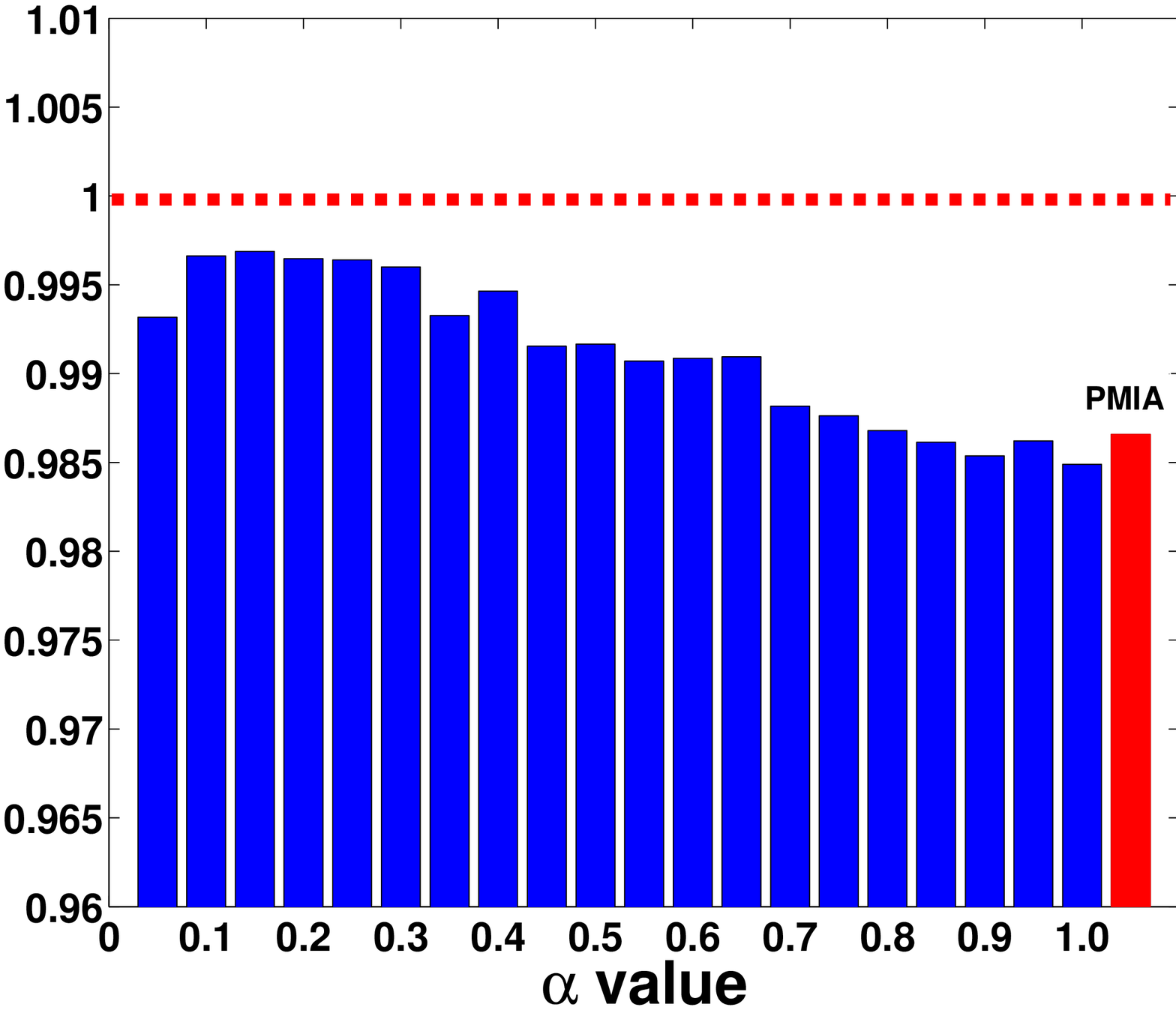}}  
    \subfigure[ \vspace{-1cm}  ca-HepPh.
    \quad]{\label{fig:PHYEffi}\includegraphics[scale=0.18]{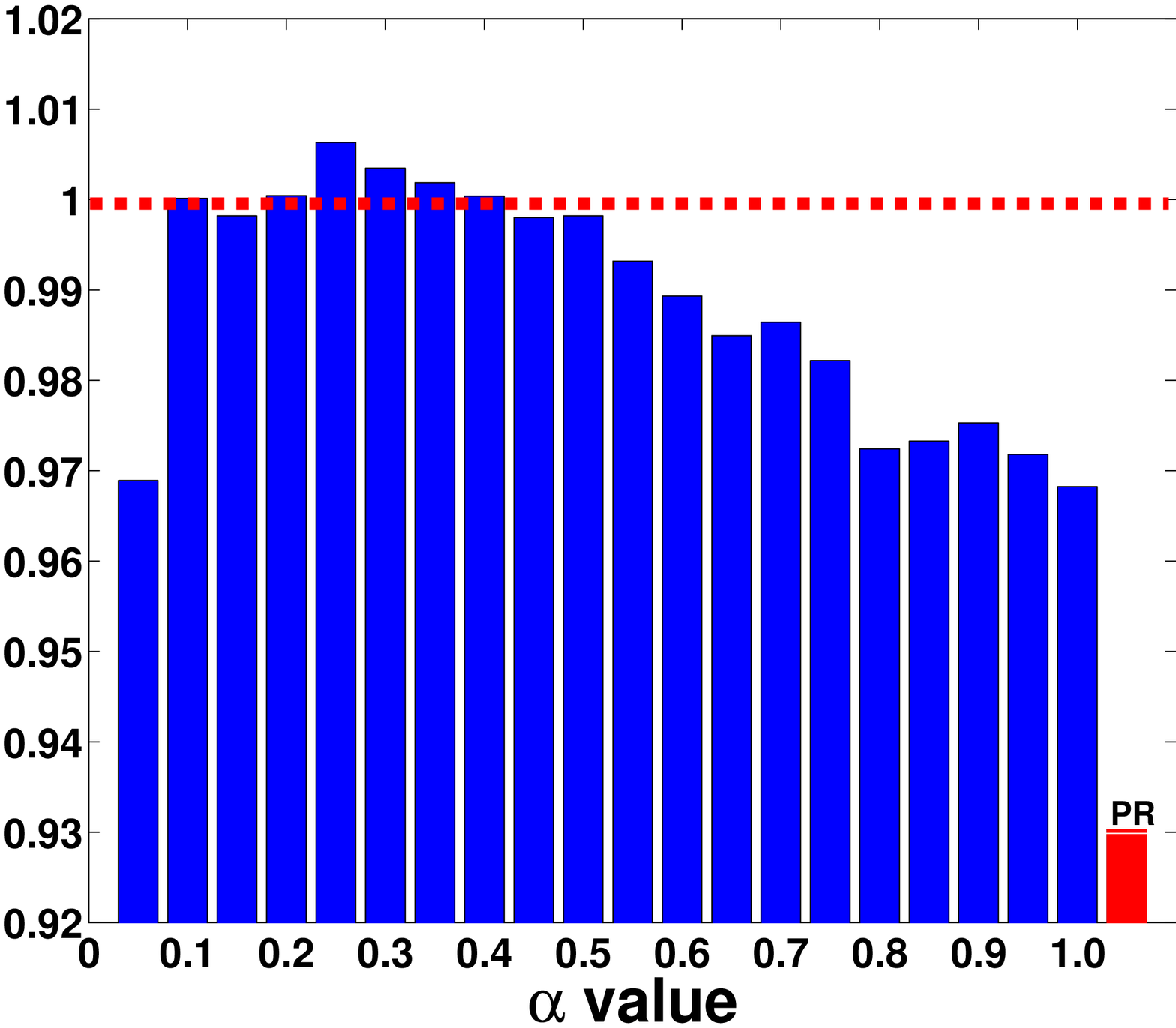}}   
    \subfigure[ \vspace{-1cm}  Wiki-Vote.
    \quad]{\label{fig:wikiTime}\includegraphics[scale=0.18]{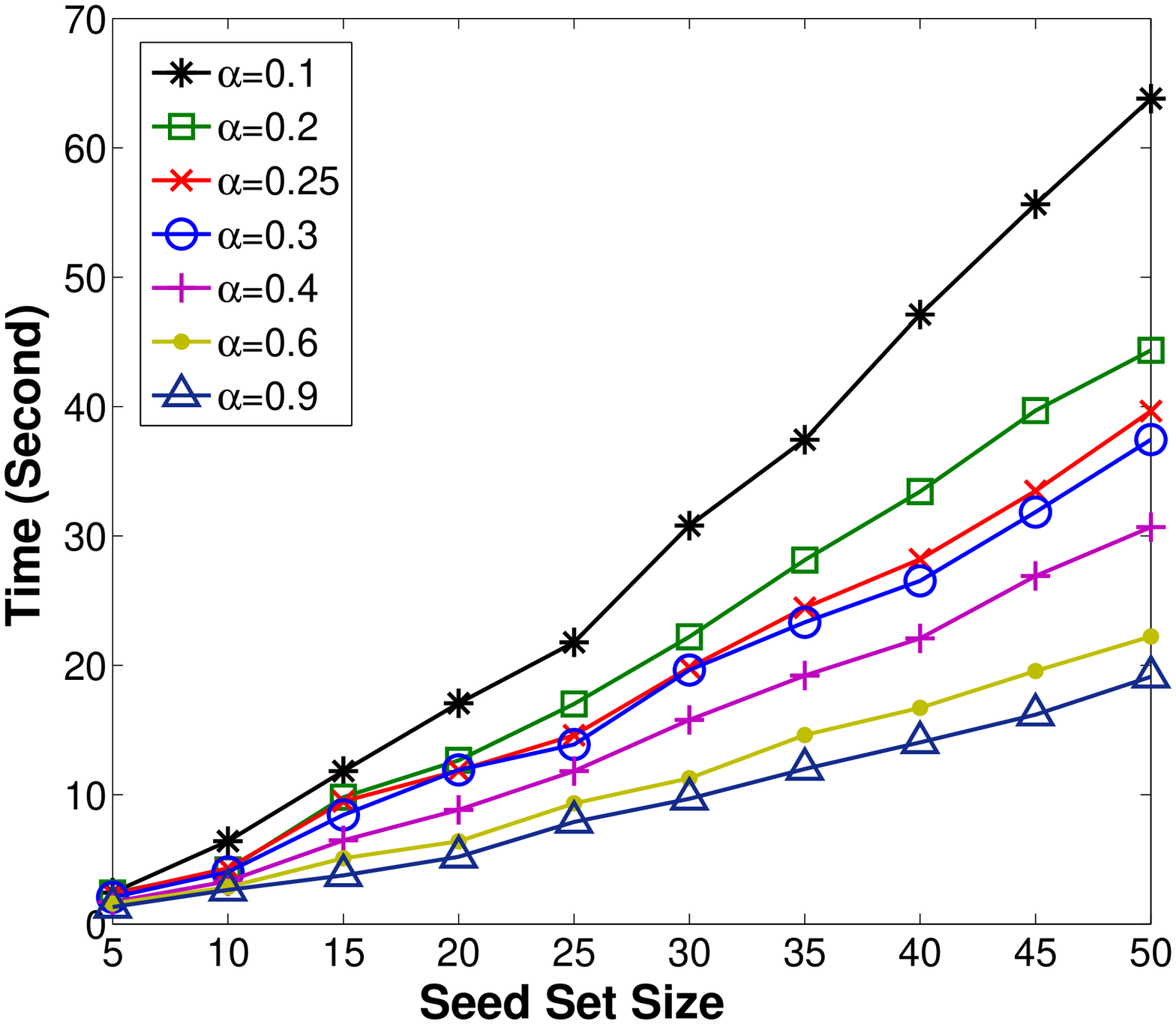}}  
    \subfigure[ \vspace{-1cm}  ca-HepPh.
    \quad]{\label{fig:PHYTime}\includegraphics[scale=0.18]{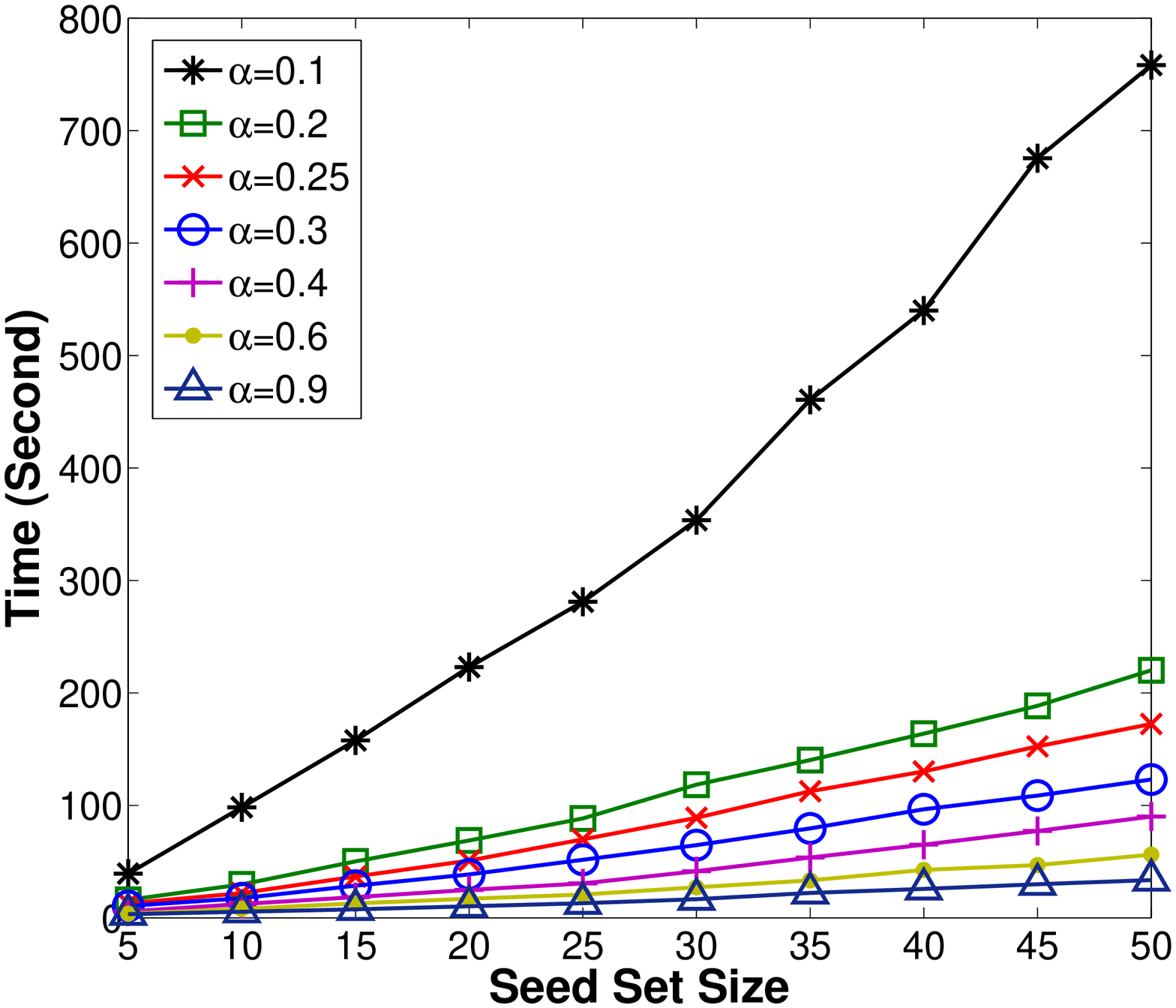}}  
  \end{center}
  \caption{The variation of effectiveness and the running time of Circuit as the change of $\lambda$ on two network datasets: Wiki-Vote,  and ca-HepPh. The top row of figures show the variation of effectiveness and the bottom row of figures show the variation of the  running time. \label{fig:lambda}}
\end{figure}

We investigate the effect of tuning parameter $\lambda$ on the
running time of Circuit and the result of its influence spread.
Specifically, we set $\lambda$ ranges from 0.05 to 1, step by 0.05,
and then get the corresponding influence spread and running time.
And, for a clear view of the influence spread results, we use the
ratio of their influence spread result relative to CELF's to
indicate their effectiveness.

Figures~\ref{fig:wikiEffi},~\ref{fig:PHYEffi}
show the effectiveness of Circuit with different $\lambda$ on Wiki-Vote,
ca-HepPh network respectively. In these figures, the x axis
is the $\lambda$ value; the red long dash line is $y=1$ which
indicates the results of CELF; the last red bar at the right side is
the effectiveness of the best algorithm among PMIA, PageRank,
DegreeDiscountIC, Degree. From these figures, we can obtain the
following observations:
\begin{itemize}
  \item The performance of of Circuit is very stable. No matter what value $\lambda$ is, the difference of effectiveness is less than 0.04, and for most of $\lambda$ values, the effectiveness of Circuit is larger than 0.97~(against CELF) and better than the best one among PMIA, PageRank, DegreeDiscountIC and Degree.
  \item When the $\lambda$ value increases from 0 to 1, the effectiveness of Circuit ascends at the beginning and then descends starting from a certain point. This observation could be explained by our assumption in Section~\ref{sec:circuit}--- there exists a certain damping coefficient in the real-world information propagation process. The father the manually set number is from the real damping value, the worse our model describes the real information propagation, and vice verse. Experimentally, the real $\lambda$ located in the range $[0.1,0.4]$.
\end{itemize}

Figures~\ref{fig:wikiTime},~\ref{fig:PHYTime}
show the running time of Circuit with different $\lambda$ (for a
better view, we show the results of $\lambda=$0.1, 0.2, 0.25, 0.3,
0.4, 0.6, 0.9 and remove the others) on Wiki-Vote, ca-HepPh
respectively.  On these figures, we can observe that the running
time of Circuit is descending with the ascending of $\lambda$. When
the $\lambda = 0.6$, the running times of Circuit are 22.2s,
56.1s respectively, which are comparable with PMIA's (10.8s,
56.3s respectively), while the influence spread results of Circuit
with $\lambda=0.6$ are all better than PMIA's. From the above
observations, we can know that if we want to get a better
effectiveness, we should set $\lambda$ to be a number in
$[0.1,0.4]$ and if we want to get the result efficiently, we should set
$\lambda$ to be a comparable large value.

\subsection{The Effectiveness of Upper Bound}

\begin{figure}
  \begin{center}
    \subfigure[ \vspace{-1cm}  Wiki-Vote.
    \quad]{\label{fig:wikiBound}\includegraphics[scale=0.18]{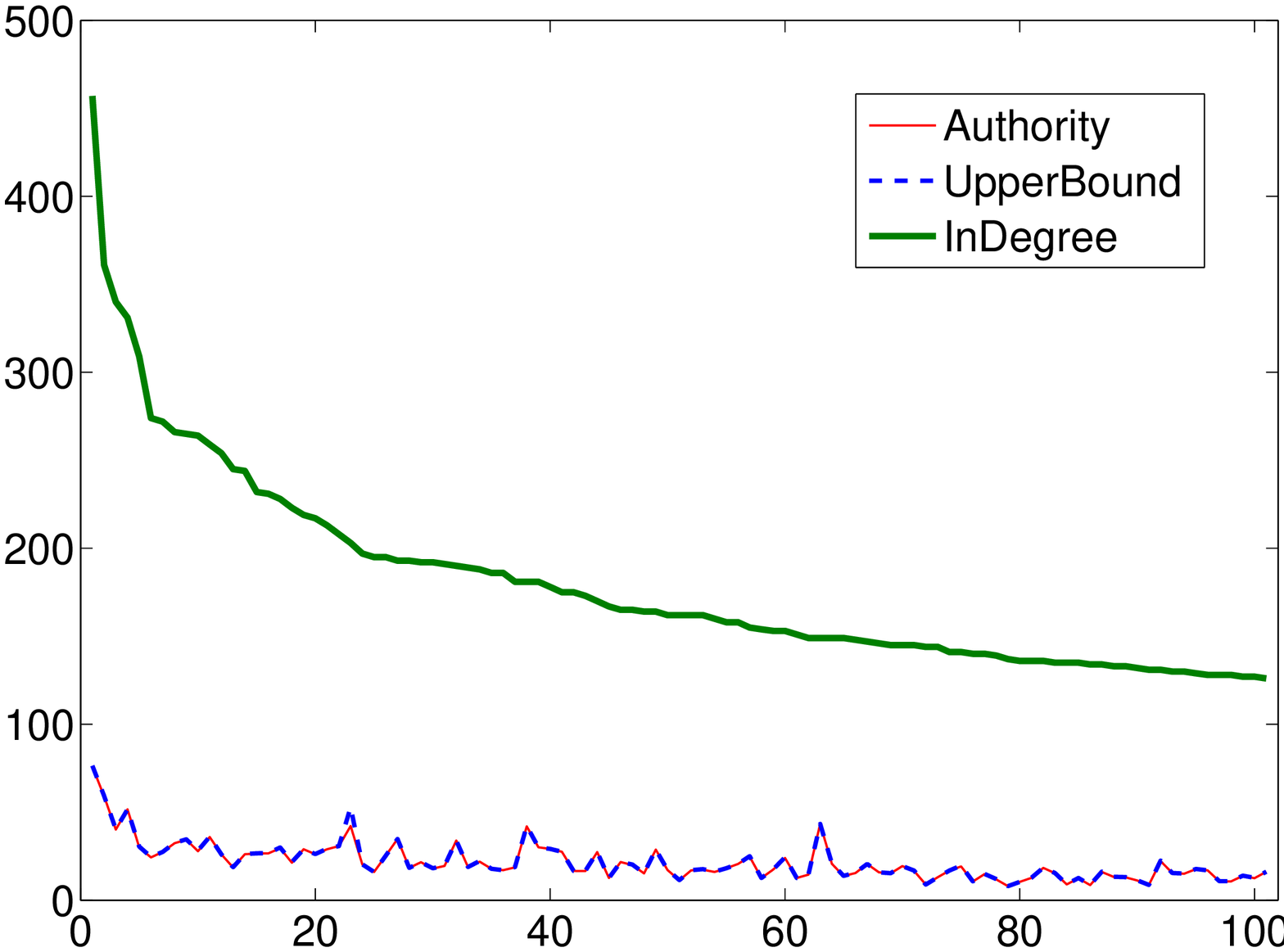}} 
    \subfigure[ \vspace{-1cm}  ca-HepPh.
    \quad]{\label{fig:PHYBound}\includegraphics[scale=0.18]{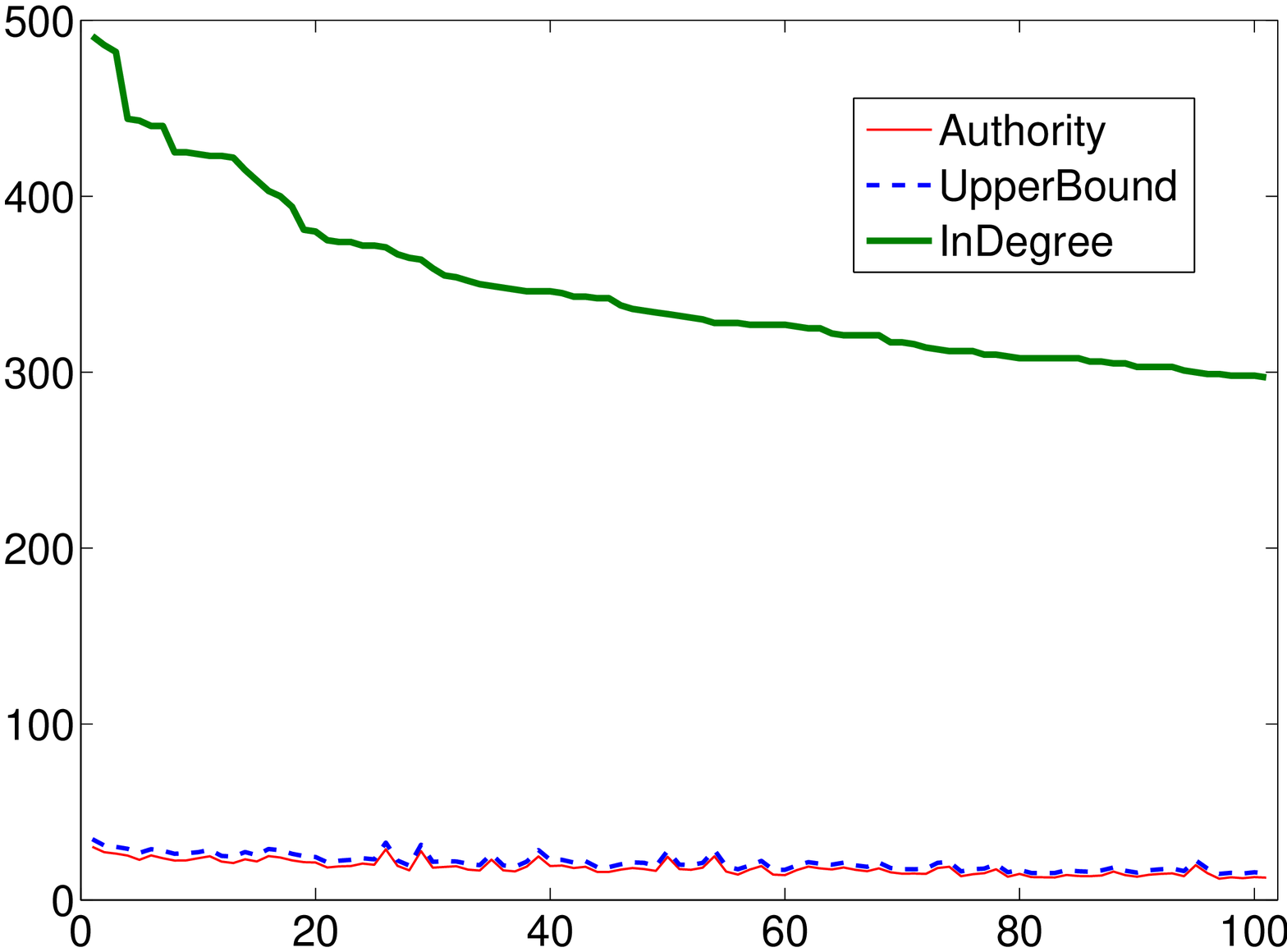}}  
  \end{center}
  \caption{The upper bound of Authority on four networks: Wiki-Vote, NetHEPT, ca-HepPh, and Amazon.  \label{fig:bound}}
\end{figure}

To demonstrate the effectiveness of the upper bound proposed in
Property~\ref{pr:potentialBound}, we take a test in the following
two networks: Wiki-Vote, ca-HepPh. We first
select top-100 nodes with the highest indegree from each network, then
compute their importance ($\mathcal{F}_i=\sum_{j=1}^{n}{f_{ij}}$) and
their corresponding upper bound $(1+\lambda)\mathcal{P}_i$(with parameter $\lambda=0.25$). We show the value of indegree, importance, upper bound of the
top-100 nodes in Figure~\ref{fig:bound}, where the green line is the
indegree value, the blue long-dash line is the upper bound value,
and the red line is the importance value. In Figure~\ref{fig:bound}
we can observe that, for each network, the blue line is always very
close to the red line which demonstrates that $(1+\lambda)\mathcal{P}_i$ is a very compact upper bound of importance $\mathcal{F}_i$. This is a very
important reason why Property~\ref{pr:potentialBound} can help us
to reduce the search space to less than 5\%~(as illustrated in
Table~\ref{table:characteristics}).

\section{Related Work}\label{sec:relatedWork}

Related work can be grouped into two categories. In the first
category, we describe some  existing social influence models.
The second category includes the existing works for the
social influence maximization problem.

\textbf{Social Influence Models.}
In the literature, many works about social influence have been published. For instance, Anagnostopoulos et
al.~\cite{anagnostopoulos2008influence} proved the existence of
social influence by statistical tests. Also, Goyal et al.~\cite{goyal2010learning}
studied how to learn the true probabilities of social influence
between individuals. In addition, there are several models to infer how the influence propagates through the network. For example, Granovetter et al.~\cite{granovetter1978threshold} proposed the Linear Threshold(LT) model to describe it, while Goldenberg et al.~\cite{goldenberg2001talk} proposed the Independent Cascade(IC) model. To the best of our knowledge, they are two most widely used models. Since these two models are not tractable, Kimura et al.~\cite{kimura2006tractable} proposed a comparably tractable model SPM  and Aggarwal et al.~\cite{aggarwalflow} proposed a stochastic model to address this issue. In addition, Tang et al.~\cite{tang2009social} proposed a Topical Affinity
Propagation approach, a graphical probabilistic model, to describe
the topic-based social influence analysis problem. Recently, Easley
et al.~\cite{easley2010networks} and Aggarwal et
al.~\cite{aggarwal2011social} summarized  and generalized many
existing studies on social influence and some other research aspects
of social networks. More importantly, they demonstrate that by
carefully study, the information exploited from social influence can
be leveraged for dealing with the real-world problems~(e.g., the
problems from markets or social security) effectively and
efficiently.

\textbf{Social Influence Maximization.}  Domingoes and Richardson proposed to exploit social influence for the marketing application, which is called viral marketing~\cite{domingos2001mining, richardson2002mining}, or social influence maximization. The goal is to find a set of seeds which will influence the maximal number of individuals in the network.

Kempe et al. formulated the influence
maximization problem as the discrete optimization problem, and they
considered three cascade models~(i.e., IC
model, Weighted Cascade(WC) model~\cite{kempe2003maximizing}, and LT model). Also, they proved that the optimization problem is NP-hard, and
presented a greedy approximation~(GA) algorithm applicable to all
three models, which guarantees that the influence spread result is within
$(1-1/e)$ of the optimal result. To address the
efficiency issue, Leskovec et al.~\cite{leskovec2007cost} presented
an optimized greedy algorithm, which is referred to as the
"Cost-Effective Lazy Forward"~(\textbf{CELF}) scheme. The CELF optimization
uses the submodularity property of the influence maximization
objective to reduce the number of evaluations on the influence
spread of nodes. To address the scalability issue, Chen et al. proposed several heuristic methods includes \textbf{DegreeDiscountIC}~\cite{chen2009efficient} and \textbf{PMIA}~\cite{chen2010scalable} which uses local
arborescence structures of each individual to approximate the social
influence propagation. Alternatively, Wang
et al.~\cite{wang2010community} presented a community-based greedy
algorithm to find the Top-$K$ influential nodes. They first detect
the communities in social network and then find influential nodes
from the selected potential communities. Among the aforementioned methods, the best algorithms in seeking the most influential seed set are those based on a Monte-Carlo
simulations to compute the influence spread for each given seed set.
The GA, CELF are all this type of
algorithms.

\section{Conclusion}\label{sec:con}
In this paper, we developed a social influence model based on circuit theory for describing the information propagation in social networks. This model is tractable and flexible for understanding patterns of information propagation. Under this model, several upper bound properties were identified. These properties can help us to quickly locate the nodes to be considered during the information propagation process. This can drastically reduce the search space, and thus vastly improve the efficiency of measuring the influence strength between any pair of nodes. In addition, the circuit theory based model provides a new way to compute the independent influence of nodes and leads to a natural solution to the social influence maximization problem. Finally, experimental results showed the advantages of the circuit theory based model over the existing models in terms of efficiency as well as the effectiveness for measuring the information propagation in social networks.

\bibliographystyle{abbrv}

\end{document}